\documentclass[twocolumn,aps,prb,a4paper,showpacs]{revtex4}
\usepackage{amsmath,amsfonts,graphicx,color}

\newcommand{\s}{\sigma}

\newcommand{\ixe}{\rangle\!\langle}
\newcommand{\phd}{\phantom{\dagger}}

\begin{document}

\title{Interference effects in the Coulomb blockade regime:\\
current blocking and spin preparation in symmetric nanojunctions}

\author{Andrea Donarini}
\affiliation{Theoretische Physik, Universit\"{a}t Regensburg,
93040 Regensburg, Germany}

\author{Georg Begemann}
\affiliation{Theoretische Physik, Universit\"{a}t Regensburg,
93040 Regensburg, Germany}

\author{Milena Grifoni}
\affiliation{Theoretische Physik, Universit\"{a}t Regensburg,
93040 Regensburg, Germany}

\date{\today}

\begin{abstract}
We consider nanojunctions in the single-electron tunnelling regime
which, due to a high degree of spatial symmetry, have a degenerate
many body spectrum. As a consequence, interference phenomena which
cause a current blocking can occur at specific values of the bias
and gate voltage. We present here a general formalism to give
necessary and sufficient conditions for interference blockade also
in the presence of spin polarized leads. As an example we analyze
a triple quantum dot single electron transistor (SET). For a
set-up with parallel polarized leads, we show how to selectively
prepare the system in each of the three states of an excited spin
triplet without application of any external magnetic field.
\end{abstract}



\maketitle

\section{Introduction}

Single particle interference is one of the most genuine quantum
mechanical effects. Since the original double-slit experiment
\cite{Young1804}, it has been observed with electrons in vacuum
\cite{Joensson61,Merli76} and even with the more massive $C_{60}$
molecules \cite{Arndt99}. Mesoscopic rings threaded by a magnetic
flux provided the solid-state analogous \cite{Yacoby95,
Gustavsson08}. Intra-molecular interference  has been recently
discussed in molecular junctions for the case of strong
\cite{Gutierrez03, Cardamone06, Ke08, Qian08} and weak
\cite{Begemann08,Darau09,Donarini09} molecule-lead coupling. What
unifies these realizations of quantum interference is that the
travelling particle has two (or more) \emph{spatially equivalent}
paths at disposal to go from one point to another of the
interferometer.

\begin{figure}[h!]
  \includegraphics[width=0.8\columnwidth,angle=0]{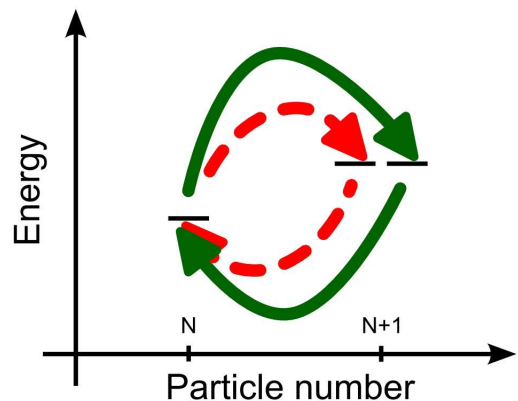}
  \caption{Interference in a single electron transistor (SET). The dynamics is governed
  by equivalent paths in the many-body spectrum that involve two (or more) degenerate states.
  }
  \label{Fig:Paths}
\end{figure}

Interference, though is hindered by decoherence. Generally, for
junctions in the strong coupling regime decoherence can be
neglected due to the short time of flight of the particle within
the interferometer. In the weak coupling case, instead, the
dwelling time is long. Usually, the decoherence introduced by the
leads dominates, in this regime, the picture and the dynamics
essentially consists of sequential tunnelling events connecting
the many-body eigenstates of the isolated system. Yet,
interference is achieved whenever two \emph{energetically
equivalent} paths involving degenerate states contribute to the
dynamics (see Fig.~\ref{Fig:Paths}). Interference survives as far
as the splitting between the many body levels is smaller that the
tunnelling rate to the leads since in this limit the system cannot
distinguish between the two paths. Thus, in such devices, that we
called interference single electron transistors\cite{Darau09}
(ISET), interference effects show up even in the Coulomb blockade
regime. They can {\it e.g.} yield a selective spin blockade in an
ISET coupled to ferromagnetic leads\cite{Donarini09}. Similar
blocking effects have been found also in multiple quantum dot
systems in dc \cite{Emary07} and ac \cite{Busl10} magnetic fields.

%
%
%
%

In the present paper we develop a general theory of interference
blockade. We give in fact an \emph{a priori} algorithm for the
detection of the interference blocking states of a generic ISET.
As a concrete example, we analyze the triple dot ISET (see
Section~\ref{sec:Triple_dot}) as it represents the simplest structure
exhibiting interference blockade. In particular we concentrate on
the blockade that involves an \emph{excited} triplet state and we
show how to prepare the system in each of the three spin states
without application of any external magnetic field. Thus we obtain
an interference mediated control of the electron spin in quantum
dots, a highly desirable property for spintronics
\cite{Wolf01,Awschalom07,Ohno00} and spin-qubit applications
\cite{Golovach06,Levitov03,Debald05,Walls07,Nowak07}.

The method of choice for the study of the dynamics in those
systems is the generalized master equation approach for the
reduced density matrix (RDM), where coherences between degenerate
states are retained \cite{Gurvitz96, Braun04, Wunsch05,
Donarini06, Hornberger08, Darau09, Donarini09, Braig05, Harbola06,
Mayrhofer07, Koller07, Pedersen07, Begemann08, Schultz09}. Such
coherences give rise to precession effects and ultimately cause
interference blockade.

The paper is organized as follows: in section \ref{sec:Model} we
introduce a generic model of ISET. In section
\ref{sec:Blocking_states} we set the necessary and sufficient
conditions which define the interference blocking states and a
generic algorithm to detect them. In section \ref{sec:Triple_dot}
we apply the theory to the triple dot molecule as archetypal
example of ISET. Section \ref{sec:Conclusions} closes the paper
with a summary of the results and conclusive remarks.

\section{Generic model of ISET}
\label{sec:Model}

Let us consider the interference single electron transistor (ISET)
described by the Hamiltonian:

\begin{equation}
  H = H_{\rm sys} + H_{\rm leads} + H_{\rm tun},
 \label{Ham_gen}
\end{equation}
where $H_{\rm sys}$ represents the central system and also
contains the energy shift operated by a capacitively coupled gate
electrode at the potential $V_{\rm g}$. The Hamiltonian $H_{\rm
sys}$ is invariant with respect to a set of point symmetry
operations that defines the symmetry group of the device. This
fact ensures the existence of \emph{degenerate} states. In
particular, for essentially planar structures belonging to the
$D_n$ group, the (non-accidental) orbital degeneracy is at maximum
twofold and can be resolved using the eigenvalues $\ell$ of the
projection of the angular momentum along the principal axis of
rotation. A generic eigenstate is then represented by the ket $|N
\ell \sigma E \rangle$ where $N$ is the number of electrons on the
system, $\sigma$ is the spin and $E$ the energy of the state. The
size of the Fock space can make the exact diagonalization of
$H_{\rm sys}$ a numerical challenge in its own. We will not treat
here this problem and concentrate instead on the transport
characteristics. $H_{\rm leads}$ describes two reservoirs of non-interacting electrons with a difference
$eV_{\rm b}$ between their electrochemical potentials. Finally,
$H_{\rm tun}$ accounts for the weak tunnelling coupling between the
system and the leads, characteristic of SETs, and we consider the
tunnelling events restricted to the atoms or to the dots closest
to the corresponding lead:

\begin{equation}
  H_{\rm tun} = t \sum_{\alpha k \sigma}
  (c^{\dagger}_{\alpha k \sigma}d^{\phd}_{\alpha \sigma}
  + d^{\dagger}_{\alpha \sigma}c^{\phd}_{\alpha k \sigma}),
\label{Ham_tun}
\end{equation}
where $c^{\dagger}_{\alpha k \sigma}$ creates an electron with
spin $\sigma$ and momentum $k$ in lead $\alpha = L,R$,
$d^{\dagger}_{\alpha \sigma}$ creates an electron in the atom or
dot closest to the lead $\alpha$ and $t$ is the bare tunnelling
amplitude that we assume for simplicity independent of $\alpha$,
$k$ and $\sigma$.

In the weak coupling regime the dynamics essentially consists of
sequential tunnelling events at the source and drain lead that
induce a flow of probability between the many-body eigenstates of
the system. It is natural to define, in this picture, a blocking state as a state which the system can enter but from which it can not escape. When the system occupies a blocking state the particle number can not change in time and the current vanishes. If degenerate states participate to transport, they can lead to interference since, like the two arms of an electronic interferometer, they are populated simultaneously. In particular, depending on the external parameters they can form linear superpositions which behave as blocking states. If a blocking state is the linear combination of degenerate states we call it \emph{interference blocking state}.

The coupling between the system and the leads not only generates
the tunneling dynamics described so far, but also contributes to
an internal dynamics of the system that leaves unchanged its particle number. In fact the equation of motion for the reduced density matrix $\rho$ of the system can be cast, to lowest non vanishing order in the coupling to the leads, in the form \cite{Braun04, Braig05, Donarini09}:
\begin{equation}
 \dot{\rho} = -\frac{i}{\hbar} [H_{\rm sys},\rho] -
\frac{i}{\hbar}[H_{\rm eff},\rho] + \mathcal{L}_{\rm tun}\rho.
\label{eq:GME-small}
\end{equation}
The commutator with $H_{\rm sys}$ in Eq.~\eqref{eq:GME-small}
represents the coherent evolution of the system in absence of the
leads. The operator $\mathcal{L}_{\rm tun}$ describes instead the
sequential tunnelling processes and is defined in terms of the
transition amplitudes between the different many-body states.
Finally, $H_{\rm eff}$ renormalizes the coherent dynamics
associated to the system Hamiltonian and is also proportional to
the system-lead tunnelling coupling. The specific form of $H_{\rm
eff}$ depends on the details of the system, yet in all cases it is
bias and gate voltage dependent and it vanishes for non degenerate
states.

\section{Blocking states}
\label{sec:Blocking_states}

\subsection{Classification of the tunnelling processes}

For the description of the tunnelling dynamics contained in the
superoperator $\mathcal{L}_{\rm tun}$ it is convenient to classify
all possible tunnelling events according to four categories:
i)Creation (Annihilation) tunnelling events that increase
(decrease) by one the number of electrons in the system, ii)
Source (Drain) tunnelling that involves the lead with the higher
(lower) chemical potential, iii) $\uparrow$ ($\downarrow$)
tunnelling that involves an electron with spin up (down) with
respect of the corresponding lead quantization axis, iv) Gain
(Loss) tunnelling that increases (decreases) the energy in the
system.

Using categories i)-iii) we can efficiently organize the matrix
elements of the system component of $H_{\rm tun}$ in the matrices:

\begin{equation}
T^{+}_{N,EE'}=
\left(%
\begin{array}{c}
  t^+_{\rm S\uparrow} \\
  t^+_{\rm S\downarrow} \\
  t^+_{\rm D\uparrow} \\
  t^+_{\rm D\downarrow} \\
\end{array}%
\right) \qquad
T^{-}_{N,EE'}=
\left(%
\begin{array}{c}
  t^-_{\rm S\uparrow} \\
  t^-_{\rm S\downarrow} \\
  t^-_{\rm D\uparrow} \\
  t^-_{\rm D\downarrow} \\
\end{array}%
\right)\\
\end{equation}
where $S,D$ means source and drain respectively and

\begin{equation}
 t^+_{\alpha \sigma} =
 \langle N+1,\{\ell',\tau'\}, E'|
 d^\dagger_{\alpha \sigma}
 |N, \{\ell,\tau\},E \rangle
 \label{eq:t_+}
\end{equation}
is a matrix in itself, defined for every creation transition from
a state with particle number $N$ and energy $E$ to one with $N+1$
particles and energy $E'$. We indicate correspondingly in the following transitions involving $t^+_{S \sigma}$ and $t^+_{D \sigma}$ as source-creation and drain-creation transitions. The compact notation $\{\ell,\tau\}$ indicates all possible combination of the quantum numbers $\ell$ and $\tau$. It follows that the size of $t^+_{\alpha \sigma}$ is ${\rm mul}(N+1,E')\times {\rm mul}(N,E)$ where the function ${\rm mul}(N,E)$ gives the degeneracy of the many-body energy level with $N$ particles and energy $E$. Analogously

\begin{equation}
 t^-_{\alpha \sigma} =
 \langle N-1,\{\ell',\tau'\}, E'|
 d^{\phd}_{\alpha \sigma}
 |N, \{\ell,\tau\},E \rangle
 \label{eq:t_-}
\end{equation}
accounts for the annihilation transitions.

\begin{figure}[h!]
  \includegraphics[width=1\columnwidth,angle=0]{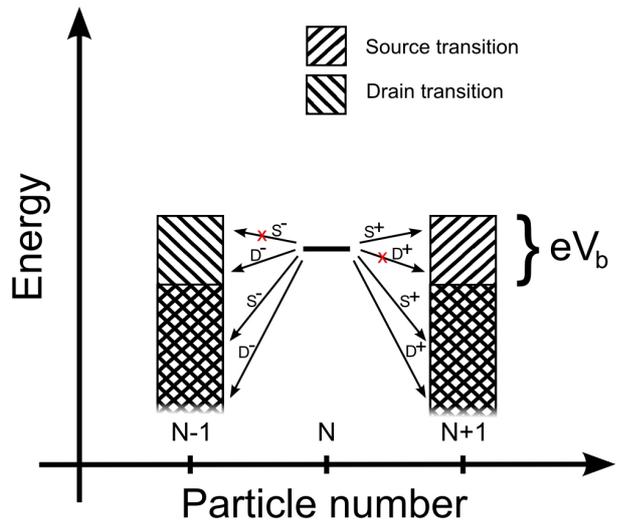}
  \caption{Energetically available transitions \emph{from} an $N$
  particle level. The patterned rectangles indicate the energy range of
  energetically available source (S) and drain (D) transitions both to states with $N+1$
  and $N-1$ particles. The arrows show examples of both allowed and forbidden transitions.
  }
  \label{Fig:Energy_cond}
\end{figure}

The fourth category concerns energy and it is intimately related
to the first and the second. Not all transitions are in fact
allowed: due to the energy conservation and the Pauli exclusion
principle holding in the fermionic leads, the energy gain (loss)
of the system associated to a gain (loss) transition is governed
by the bias voltage. These energy conditions are summarized in the
table \ref{tab:En_condition} and illustrated in
Fig.~\ref{Fig:Energy_cond}.

\begin{table}[h!]
\begin{tabular}{c|c|c}

  $\Delta E \lesssim$ & Creation & Annihilation \\
  \hline
   Source  & $+eV_{\rm b}/2$ & $-eV_{\rm b}/2$ \\
  \hline
   Drain & $-eV_{\rm b}/2$ & $+eV_{\rm b}/2$ \\

\end{tabular}
\caption{Energy conditions for tunnelling transitions between the
many-body eigenstates of the system. The quantity $\Delta E = E_f
- E_i$ is the difference between the energies of the final and
initial many-body states of the system involved in the
transition.} \label{tab:En_condition}
\end{table}
The quantity $\Delta E := E_f - E_i$, is the difference between
the energy of the final and initial state of the system and the
approximate condition $\lesssim$ is due to the thermal broadening
of the Fermi distributions. For simplicity we set the zero of the
energy at the chemical potential of the unbiased device and we
assume an equal potential drop at the source and drain contact. In
the table \ref{tab:En_condition} one reads for example that in a
source-creation tunnelling event the system can gain at maximum
$\tfrac{eV_{\rm b}}{2}$ or that in a source-annihilation and
drain-creation transition the system looses at least an energy of
$\tfrac{eV_{\rm b}}{2}$.

From table \ref{tab:En_condition} one also deduces that, from
whatever initial state, it is always possible to reach the lowest
energy state (the global minimum) via a series of energetically
allowed transitions. Vice versa, not all states can be reached
starting from the global minimum. Thus, the only \emph{relevant
states} for the transport in the stationary regime are the states
that can be reached from the global minimum via a finite number of
energetically allowed transitions.

%
%
%
%

\subsection{Subspace of decoupled states}

In the process of detecting the blocking states we observe first
that some states do not participate to the transport and can be
excluded {\it a priori} from any consideration. These are states
with zero transition elements to all other \emph{relevant} states.
Within the subspace with $N$ particles and energy $E$ the
decoupled states span the vector space:

\begin{equation}
\mathcal{D}_{N,E} = \bigcap_{E'} \left[
 {\rm ker}\,\, T^+_{N,EE'}\cap {\rm ker}\,\, T^-_{N,EE'}
 \right]
\label{eq:Decoupled_states}
\end{equation}
where $E'$ is the energy of a relevant state with $N+1$ or $N-1$
particles respectively. The function ${\rm ker}\,\,M$ returns the
null space of the linear application associated to the matrix $M$.

The decoupled space $\mathcal{D}_{N,E}$ as presented in equation
\eqref{eq:Decoupled_states} is constructed as follows. Let us
consider a generic many-body state $|\psi_{NE}\rangle$ with $N$
particles and energy $E$ and let ${\bf v}$ be the vector of its
components in the basis $| N \ell \tau E \rangle$. The vector
$T^+_{N,EE'} \mathbf{v}$ has thus $4\times{\rm mul}(N+1, E')$
components and consists of all possible transition amplitudes from
$|\psi_{NE}\rangle$ to all possible states with $N+1$ particles
and energy $E'$. Consequently ${\rm ker}\,\, T^+_{N,EE'}$ contains
the vectors ${\bf v}$ associated to states with $N$ particles and
energy $E$ which are decoupled from all possible states with $N+1$
particles and energy $E'$. Analogously holds for the significance
of ${\rm ker}\,\, T^-_{N,EE'}$. The intersections in
\eqref{eq:Decoupled_states} and the condition on $E'$ ensure that
$\mathcal{D}_{N,E}$ contains only states decoupled at the same
time from all other states relevant for transport in the
stationary regime. We emphasize that, due to the condition on the
energy $E'$, the decoupled space $\mathcal{D}_{N,E}$ is a
dynamical concept that depends on the applied gate and bias across
the ISET. The coupled space $\mathcal{C}_{N,E}$ is the orthogonal
complement of $\mathcal{D}_{N,E}$ in the Hilbert space with $N$
particles and energy $E$. The blocking states belong to it.

As a first simple application of the ideas presented so far, let
us consider the SET at zero bias. According to the table
\ref{tab:En_condition} the system can only undergo loss tunnelling
events and the global energy minimum is the only blocking state,
in accordance with the observation that the system is in
equilibrium with the leads and that we measure the energy starting
from the equilibrium chemical potential\cite{Free_energy}. The
potential $V_{\rm g}$ of the gate electrode defines the particle
number of the global minimum and, by sweeping $V_{\rm g}$ at zero
bias, one can change the number of electrons on the system one by
one. This situation, the Coulomb blockade, remains unchanged until
the bias is high enough to open a gain transition that unblocks
the global minimum. Then, the current can flow. Depending on the
gate this first unblocking transition can be of the kind source-creation or drain-annihilation. Correspondingly, the current is associated
to $N \leftrightarrow N+1$ or $N \leftrightarrow N-1$
oscillations, where $N$ is the particle number of the global
minimum.

\subsection{Blocking conditions}

At finite bias the condition which defines a blocking state
becomes more elaborate:

\begin{enumerate}
    \item The blocking state must be achievable from the global
    minimum with a finite number of allowed transitions.
    \item All matrix elements corresponding to energetically
    allowed transitions outgoing from the blocking state
    should vanish: in particular all matrix elements corresponding to $E_f - E_{\rm block} <
    -\tfrac{eV_b}{2}$ and for $|E_f - E_{\rm block}| < \tfrac{eV_b}{2}$ only
    the ones corresponding to the drain-annihilation and source-creation transitions.
\end{enumerate}

The first condition ensures the blocking state to be populated in
the stationary regime. The second is a modification of the generic
definition of blocking state restricted to energetically allowed
transitions and it can be written in terms of the tunnelling
matrices $T^+_{N,EE'}$ and $T^-_{N,EE'}$. For each many-body
energy level $|NE\rangle$, the space spanned by the blocking
states reads then:

\begin{equation}
\mathcal{B}_{N,E} = \mathcal{B}^{(1)}_{N,E} \cap
\mathcal{B}^{(2)}_{N,E} \cap \mathcal{C}_{N,E}
\label{eq:Blocking_space}
\end{equation}
with
\begin{equation}
\begin{split}
\mathcal{B}^{(1)}_{N,E} = \bigcap_{E'}
 &\left\{
 \mathcal{P}_{NE}\left[{\rm ker}\,\,(T^+_{N,EE'},T_D)\right] \cap \right.\\
 & \left. \mathcal{P}_{NE}\left[{\rm ker}\,\,(T^-_{N,EE'},T_S)\right]
 \right\}\\
\mathcal{B}^{(2)}_{N,E} = \bigcap_{E''}
 &\left[
 {\rm ker}\,\,T^+_{N,EE''}\cap {\rm ker}\,\,T^-_{N,EE''}
 \right].
\end{split}
\label{eq:Blocking_conditions}
\end{equation}
In Eq.~\ref{eq:Blocking_conditions} we introduced the matrices
$T_D = (\mathbf{0},\mathbf{1})^T$ and $T_S =
(\mathbf{1},\mathbf{0})^T$ with $\mathbf{1}$ being the identity
matrix and $\mathbf{0}$ the zero matrix, both of dimension
$2\times {\rm mul}(N + 1,E')$ for $T_D$ and $2\times {\rm mul}(N -
1,E')$ for $T_S$. The energies $E'$ and $E''$ satisfy the
inequalities $|E' - E|<\tfrac{eV_{\rm b}}{2} $ and $E''-E <
-\tfrac{eV_{\rm b}}{2}$, respectively, and $\mathcal{P}_{NE}$ is
the projection on the $N$ particle space with energy $E$.

The first kernel in $\mathcal{B}^{(1)}_{N,E}$ together with the
projector $\mathcal{P}_{NE}$ gives all linear combinations of $N$
particle degenerate states which have a finite creation transition
involving the drain but not the source lead. This condition can in
fact be expressed as a non-homogeneous linear equation for the
vector $\mathbf{v}$ of the components in the many body basis of
the generic $N$ particle state with energy $E$:

\begin{equation}
 T^+_{N,EE'}\mathbf{v} = \mathbf{b},
 \label{eq:Blocking_simple}
\end{equation}
where $\mathbf{b}$ is a generic vector of length $4 \times {\rm
mul}(N+1,E')$ whose first $2 \times {\rm mul}(N+1,E')$ components
(the source transition amplitudes) are set to zero. Due to the
form of $\mathbf{b}$, it is convenient to transform Eq.~
\eqref{eq:Blocking_simple} into an homogeneous equation for a
larger space of dimension ${\rm mul}(N,E) + 2 \times {\rm
mul}(N+1,E')$ which also contains the non-zero elements of
$\mathbf{b}$ and finally project the solutions of this equation on
the original space. With this procedure we can identify the space
of the solutions of \eqref{eq:Blocking_simple} with:

\begin{equation}
V = \mathcal{P}_{NE}\left[{\rm ker}\,\,(T^+_{N,EE'},T_D)\right].
\end{equation}

The second kernel in $\mathcal{B}^{(1)}_{N,E}$ takes care of the
annihilation transitions in a similar way. Notice that $V$ also
contains vectors that are decoupled at both leads. This redundance
is cured in \eqref{eq:Blocking_space} by the intersection with the
coupled space $\mathcal{C}_{NE}$.

The conditions \eqref{eq:Blocking_conditions} are the
generalization of the conditions over the tunnelling amplitudes
that we gave in [\onlinecite{Darau09}]. That very simple condition
captures the essence of the effect, but it is only valid under
certain conditions: the spin channels should be independent, the
relevant energy levels only two and the transition has to be
between a non degenerate and a doubly degenerate level. Equation
\eqref{eq:Blocking_conditions}, on the contrary, is completely
general. In appendix \ref{app:Equivalence} we give an explicit
derivation of the equivalence of the two approaches in the simple
case.

For most particle numbers $N$ and energies $E$, and sufficiently
high bias, $\mathcal{B}_{N,E}$ is empty. Yet, blocking states
exist and the dimension of $\mathcal{B}_{N,E}$ can even be larger
than one as we have already proven for the benzene and the triple
dot ISETs \cite{Donarini09}. Moreover, it is most probable to find
interference blocking states among ground states due to the small
number of intersections appearing in
\eqref{eq:Blocking_conditions} in this situation. Nevertheless
also excited states can block the current as we will show in the
next section.

The case of spin polarized leads is already included in the formalism both in the parallel and non parallel configuration.
In the parallel case one quantization axis is naturally defined on the all structure and $\sigma$ in equations \eqref{eq:t_+} and \eqref{eq:t_-} is defined along this axis.
In the case of non parallel polarized leads instead it is enough to consider $d^{\dagger}_{\alpha \sigma}$ and $d^{\phd}_{\alpha \sigma}$ in equations \eqref{eq:t_+} and \eqref{eq:t_-}, respectively, with
$\sigma$ along the quantization axis of the lead $\alpha$. It is
interesting to note that in that case, no blocking states can be
found unless the polarization of one of the leads is $P = 1$. The
spin channel can in fact be closed only one at the time via linear
combination of different spin states.

A last comment on the definition of the blocking conditions is
necessary. A blocking state is a stationary solution of the
equation \eqref{eq:GME-small} since by definition it does not
evolve in time. The density matrix associated to one of the
blocking states discussed so far i) commutes with the system
Hamiltonian since it is a state with given particle number and
energy; ii) it is the solution of the equation $\mathcal{L}_{\rm
tun}\rho = 0$ since the probability of tunnelling out from a
blocking state vanishes, independent of the final state.
Nevertheless, a third condition is needed to satisfy the
condition of stationarity:
\begin{enumerate}
 \item[3.] The density matrix $\rho_{\rm block}$ associated to the blocking state
 should commute with the effective Hamiltonian $H_{\rm eff}$ which
 renormalizes the coherent dynamics of the system to the lowest non
 vanishing order in the coupling to the leads:
 \begin{equation}
 [\rho_{\rm block},H_{\rm eff}] = 0.
 \label{eq:Condition_3}
\end{equation}
\end{enumerate}

The specific form of $H_{\rm eff}$ varies with the details of the
system. Yet its generic bias and gate voltage dependence implies
that, if present, the current blocking occurs only at specific
values of the bias for each gate voltage. Further, if an energy
level has multiple blocking states and the effective Hamiltonian
distinguishes between them, selective current blocking, and
correspondingly all electrical preparation of the system in one
specific degenerate state, can be achieved. In particular, for spin polarized leads, the system can be prepared in a particular spin state without the application of any external magnetic field as we will show explicitly in section \ref{sec:TD-c}.

\section{The triple dot ISET}
\label{sec:Triple_dot}
\begin{figure}[h!]
  \includegraphics[width=1\columnwidth,angle=0]{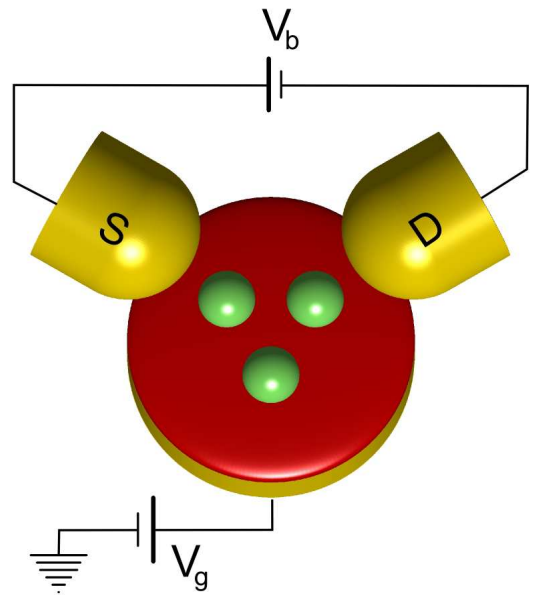}
  \caption{Schematic representation of a triple dot interference single electron transistor (ISET).
  }
  \label{Fig:System}
\end{figure}

The triple dot SET has been recently in the focus of intense theoretical\cite{Emary07,Delgado08,Gong08,Kostyrko09,Shim09,Poeltl09,Busl10} and experimental\cite{Gaudreau06,Rogge08,Gaudreau09,Austing10} investigation due to its capability of combining incoherent transport characteristics and signatures of molecular coherence.
The triple dot ISET that we consider here (Fig.~\ref{Fig:System}) is the simplest structure with symmetry protected orbital degeneracy exhibiting interference blockade. Despite its relative simplicity this system displays different kinds of current blocking and it represents for this reason a suitable playground for the ideas presented so far.

\subsection{The model}

We describe the system with an Hamiltonian in the extended Hubbard
form:

\begin{equation}
    \begin{split}
    H_{\rm sys} = \,\,
    &\xi_0 \sum_{i\s} d^{\dagger}_{i\s}d^{\phd}_{i\s} +
    b  \sum_{i\s}\left(d^{\dagger}_{i\s}d^{\phd}_{i+1\s} +
d^{\dagger}_{i+1\s}d^{\phd}_{i\s}\right)\\
    +& U \sum_i \left(n_{i\uparrow} - \tfrac{1}{2}\right)
                \left(n_{i\downarrow} - \tfrac{1}{2}\right)\\
    +& V \sum_i\left(n_{i\uparrow} + n_{i\downarrow}- 1\right)
               \left(n_{i+1\uparrow} + n_{i+1\downarrow} -
               1\right),
    \end{split}
    \label{eq:PPP}
\end{equation}
where $d^{\dagger}_{i\sigma}$ creates an electron of spin $\sigma$
in the ground state of the quantum dot $i$. Here $i =
1,\ldots,\,3$ runs over the three quantum dots of the system and
we impose the periodic condition $d_{4\sigma} = d_{1\sigma}$.
Moreover $n_{i\sigma} = d^{\dagger}_{i\sigma} d^{\phd}_{i\sigma}$.
The effect of the gate is included as a renormalization of the
on-site energy $\xi = \xi_0 - eV_{\rm g}$ where $V_{\rm g}$ is the
gate voltage. We measure the energies in units of the modulus of
the (negative) hopping integral $b$. The parameters that we use
are $\xi_0 = 0,\, U= 5\,|b|,\, V= 2\,|b|.$

%
%
%
%

\begin{figure}[h!]
  \includegraphics[width=0.9\columnwidth,angle=0]{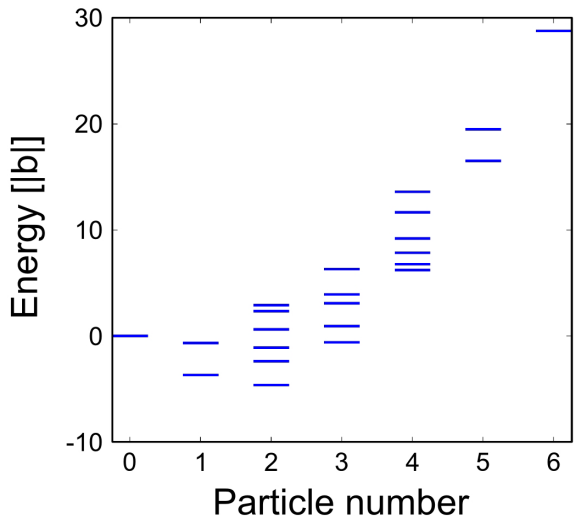}
  \caption{Spectrum of the triple dot system for the specific gate
  voltage $eV_g = 4.8 b$ chosen to favor a configuration with two electrons.
  The other parameters in the system are $U = 5|b|$ and $V =
  2|b|$, where $b$ is the hopping integral between the different
  dots.
  }
  \label{Fig:Spectrum}
\end{figure}

The number of electrons considered for the
triple dot structure goes from 0 to 6. Thus the entire Fock
space of the system contains $4^3 = 64$ states. By
exact diagonalization we obtain the many body-eigenstates and the
corresponding eigenvalues that we present in
Fig.~\ref{Fig:Spectrum} for a gate voltage of $V_{\rm g} =
4.8\,b/e$. In the table \ref{tab:Degeneracies} we also give the
degeneracies of all levels relevant for the blocking states
analysis which will follow. We distinguish between spin and
orbital degeneracy since the latter is the most important for the
identification of the blocking states. The total degeneracy of a
level is simply the product of the two.

\begin{table}[h!]
\begin{tabular}{c|c|c}

  Many-body  & Orbital & Spin \\
  energy level  & degeneracy & degeneracy \\
  \hline
  $0$    & 1 & 1 \\
  $1_0$  & 1 & 2 \\
  $2_0$  & 1 & 1 \\
  $2_1$  & 2 & 3 \\
  $3_0$  & 2 & 2 \\
  $4_0$  & 1 & 3 \\
  $5_0$  & 2 & 2 \\
  $6$    & 1 & 1 \\

\end{tabular}
\caption{Degeneracy of the triple dot system energy levels as it
follows from the underlying $D_3$ symmetry. A level $N_i$ is the
$i$th excited level with $N$ particles. The total degeneracy of
the level is the product of its orbital and spin degeneracies.}
\label{tab:Degeneracies}
\end{table}

%
%
%
%

\subsection{Excited state blocking}

\begin{figure}[h!]
  \includegraphics[width=\columnwidth,angle=0]{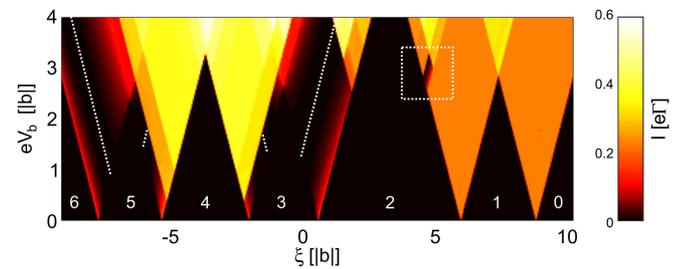}
  \caption{Stationary current for the triple dot ISET. Coulomb
  blockade diamonds are visible at low biases. Ground state and excited state
  interference blockades are also highlighted. The temperature is $k_bT = 0.002
  |b|$. The other parameters are the same as the ones in Fig.~\ref{Fig:Spectrum}.
  }
  \label{Fig:Current}
\end{figure}

In Fig.~\ref{Fig:Current} we show the stationary current through
the triple dot ISET as a function of bias and gate voltage. At low
bias the current vanishes almost everywhere due to Coulomb
blockade. The particle number is fixed within each Coulomb diamond
by the gate voltage and the zero particle diamond is the first to
the right. The zero current lines running parallel to the borders
of the 6, 4 and 2 particle diamonds are instead signatures of
ground state interference that involves an orbitally non-degenerate
ground state (with 2, 4, and 6 particle) and an orbitally
double-degenerate one (with 3 and 5 particles). In appendix
\ref{app:Equivalence} we illustrate how to obtain an expression
for the blocking states in these cases.

\begin{figure}[ht]
  \includegraphics[width=0.9\columnwidth,angle=0]{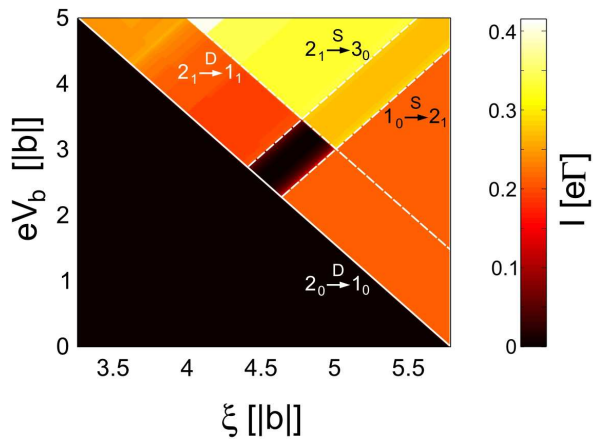}
  \caption{Blow up of the stationary current through the triple dot ISET
  around the 2 to 1 particle degeneracy point. The black area
  sticking out of the 2 particles Coulomb diamond denotes the excited states blocking.
  }
  \label{Fig:Stab_diagr}
\end{figure}

The striking feature in Fig.~\ref{Fig:Current} is the black area
of current blocking sticking out of the right side of the two
particles Coulomb diamond. It is the fingerprint of the occupation
of an excited interference blocking state.
Fig.~\ref{Fig:Stab_diagr} is a zoom of the current plot in the
vicinity of this excited state blocking. The dashed lines indicate
at which bias and gate voltage a specific transition is
energetically allowed, with the notation $N_i$ labelling the $i$th
excited many-body level with $N$ particles. These lines are
physically recognizable as abrupt changes in the current and run
all parallel to two fundamental directions determined by the
ground state transitions. For positive bias, positive (negative)
slope lines indicates the bias threshold for the opening of source-creation (drain-annihilation) transitions. The higher the bias the
more transitions are open, the higher, in general, the current.


%
%
%
%


The anomalous blockade region is delimited on three sides by
transitions lines associated to the first excited two particle
level $2_1$. Our group theoretical analysis shows that the two
particle first excited state is a twofold orbitally degenerate
spin triplet (see Table \ref{tab:Degeneracies}). In other terms we
can classify its six states with the notation $|2_1,\ell,S\rangle$
with $\ell = \pm \hbar$ being the projection of the angular
momentum along the main rotation axis, perpendicular to the plane
of the triple dot, and $S_z = -\hbar,0,\hbar$ the component of the
spin along a generic quantization axis. The $1_0$ energy level is
instead twice spin degenerate and invariant under the symmetry
operations of the point group $D_3$.

%
%
%
%

In order to identify the 2 particle blocking states we perform the
analysis presented in the previous section for the $2_1$ energy
level with the gate and bias in the blocking region. Firstly, we
find that the $2_1$ energy level can be reached from $2_0$ via the
drain-annihilation transition $2_0 \rightarrow 1_0$ followed by
the source-creation transition $1_0 \rightarrow 2_1$. Secondly,
the space of the decoupled states $\mathcal{D}_{2_1}$ is empty and
the only energetically allowed outgoing transition is the
drain-annihilation $2_1 \rightarrow 1_0$ transition. Thus the
blocking space is given by the expression:

\begin{equation}
\mathcal{B}_{2_1} = \mathcal{P}_{2_1} \left[{\rm ker}
\,(T^-_{2,2_1\,1_0}, T_S) \right]
\end{equation}
and has dimension three. For clearness we give in the appendix
\ref{app:T_matrices} the explicit expression of $T^-_{2,2_1\,1_0}$
and the corresponding vectors that span $\mathcal{B}_{2_1}$.
Essentially, there is a blocking state for each of the three
projection of the spin $S_z$. This result is natural since, for
unpolarized or parallel polarized leads, coherences between states
of different spin projection along the common lead quantization
axis do not survive in the stationary limit.

Outside the blocking region either the first or the second
blocking state conditions are violated. In particular, below the
lower right border the state $2_1$ can not be reached from the
global minimum since the $1_0 \rightarrow 2_1$ source-creation
transition is forbidden while above the upper left (right) borders
the state $2_1$ can be depopulated towards the $3_0$ ($1_0$)
states via a source-creation (drain-annihilation) transition.

\subsection{Spin polarized transport}
\label{sec:TD-c}
The orbital interference blocking presented in the previous
section acquires a spin dependence in presence of polarized leads.
The lead polarization $P_\alpha$ with $\alpha = L,R$ is defined by means of the density of states $D_{\alpha \sigma}$ at the Fermi energy for the different spin states:

\begin{equation}
P_\alpha = \frac{D_{\alpha \uparrow} - D_{\alpha \downarrow}}{D_{\alpha \uparrow} + D_{\alpha \downarrow}}
\end{equation}

and is taken equal for the two leads $P = P_L = P_R$. Finally, the spin polarization influences the dynamics of the system via the spin dependent bare tunnelling rates $\Gamma^0_{\alpha \sigma} = \tfrac{2\pi}{\hbar}|t|^2 D_{\alpha \sigma}$ that enter the definition of the tunnelling component of the Liouvillian $\mathcal{L}_{\rm tun}$ and the renormalization frequencies $\omega_{\alpha S_z}$.
We assume the leads to be parallel polarized so that no spin
torque is active in the device and we can exclude the spin
accumulation associated to that\cite{Braun04,Braig05}.

\begin{figure}[h!]
  \includegraphics[width=0.9\columnwidth,angle=0]{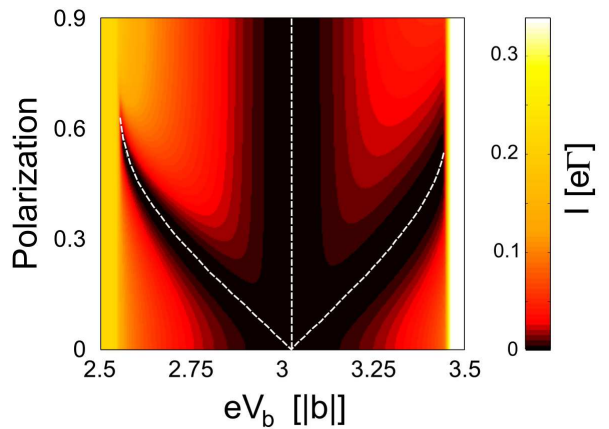}
  \caption{Current as a function of the polarization and of the
  bias voltage in proximity of the excited state interference blocking.
  The white dashed lines are defined by the conditions
  $\omega_{SS_z}=0$ for $S_z = -\hbar,0,\hbar$ from left to right,
  respectively.
  }
  \label{Fig:I_vs_PVb}
\end{figure}

\begin{figure}[h!]
  \includegraphics[width=0.9\columnwidth,angle=0]{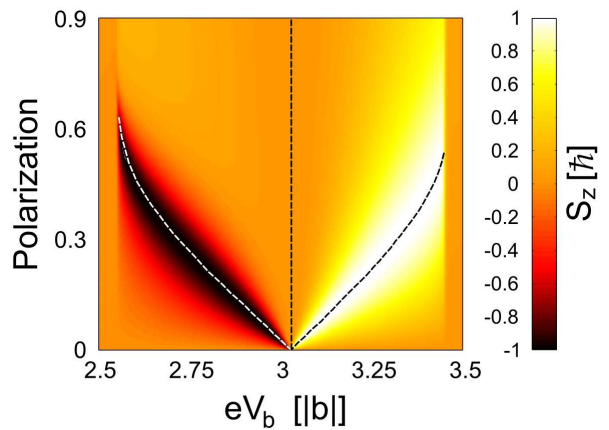}
  \caption{Spin projection $S_z$ as a function of the polarization and of the
  bias in proximity of the excited state interference blocking.
  Notice by comparison with Fig.~\ref{Fig:I_vs_PVb} the
  correspondence between current blocking and spin preparation.
  }
  \label{Fig:Sz_vs_PVb}
\end{figure}

In Fig.~\ref{Fig:I_vs_PVb} we show the current in the excited
state blocking region as a function of the bias and of the
(parallel) lead polarization $P$. For non-polarized leads the
current is blocked at a single bias, while for finite values of
$P$ the blocking is threefold. For the same bias and polarization
ranges we present in Fig.~\ref{Fig:Sz_vs_PVb} the $z$ component of
the spin for the triple dot. The spin projection $S_z$ assumes,
exactly in correspondence of the current blocking, the values $S_z
= -\hbar, 0, \hbar$, respectively, as the bias is increased.

%
%
%
%

%
%
\begin{figure}[h!]
  \includegraphics[width=0.9\columnwidth,angle=0]{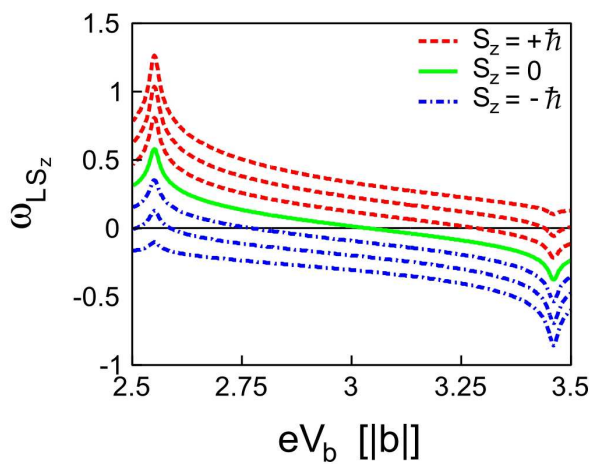}
  \caption{Renormalization frequencies as a function of the bias
  for different polarizations $P = 0, 0.25, 0.5, 0.75$ in the leads. The gate is $V_g = 4.8 \,b/e$.
  At $P=0$ all the renormalization frequencies coincide (full line). The $S_z =
  \hbar(-\hbar)$ frequency increases (decreases) monotonously with the polarization. The $S_z = 0$
  frequency is instead independent of it.
  }
  \label{Fig:wLs}
\end{figure}

%
%
The explanation of this effect relies on the third blocking
condition, Eq.~\eqref{eq:Condition_3}, and concerns the form of
the effective Hamiltonian introduced in Eq.~\eqref{eq:GME-small}.
The latter can be written, due to the rotational symmetry of the
system, in the form:

\begin{equation}
H_{\rm eff} = \sum_{\alpha S_z} \omega_{\alpha S_z} L_{\alpha},
\label{eq:H_eff}
\end{equation}
where $L_{\alpha}$ is the projection of the angular momentum in
the direction of the lead $\alpha$ and it does not depend on the
spin degree of freedom $S_z$. Moreover, $\omega_{\alpha S_z}$ is
the frequency renormalization given to the states of spin
projection $S_z$ by their coupling to the $\alpha$ lead. In the
appendix \ref{app:Effective Hamiltonian} we give an explicit
expression for $\omega_{\alpha S_z}$ and $L_{\alpha}$. In
Fig.~\ref{Fig:wLs} we plot instead $\omega_{L S_z}$ as a function
of the bias for different polarizations. The gate is fixed at $V_g
= 4.8 b/e$.

%
%
%
%

Since the two particle ground state is totally symmetric ($A_{1}$
symmetry), a three particle blocking state must be antisymmetric
with respect to the vertical plane that intersects the center of
the system and the drain dot. For this reason a blocking state is
also an eigenstate of the projection $L_{\rm D}$ of the angular
momentum in the direction of the drain lead. Consequently, the
last blocking condition is satisfied only if:
\begin{equation}
   \omega_{S S_z} = 0
   \label{eq:third_condition}
\end{equation}
and the effective Hamiltonian is proportional to $L_{\rm D}$.

For zero polarization in the leads the condition
\eqref{eq:third_condition} holds at the same bias for the three
spin projections $S_z$ and the blocking state is a statistical
mixture of the three spin projections. For polarized leads,
instead, each spin projection is blocked at a specific bias and
the spin on the system is controlled simply by changing the bias
across the device. The dashed lines in Figs.
\ref{Fig:I_vs_PVb} and \ref{Fig:Sz_vs_PVb} represent the solutions
of equations \eqref{eq:third_condition} for
$S_z=-\hbar,\,0,\,\hbar$ from left to right, respectively. Clearly
they also indicate in Fig.~\ref{Fig:I_vs_PVb} the zeros of the
current and in Fig.~\ref{Fig:Sz_vs_PVb} the fully populated spin
states.

\section{Conclusions}
\label{sec:Conclusions}

In this paper we addressed the interference effects that
characterize the transport through a symmetric single electron
transistor. In particular we gave the generic conditions for
interference blockade and an algorithm for the identification of
the interference blocking states as linear combination of
degenerate many-body eigenstates of the system.

As an application of the theory we studied the triple dot ISET.
Despite its relative simplicity, this system exhibits different
types of interference blocking and it represents an interesting
playground of the general theory. Specifically, we concentrated on
the interference blockade that involves an excited triplet state.
In presence of polarized leads we exploited the interference
blocking in order to access each of the triplet states by all
electrical means.

The theory is sufficiently general to be applied to any device
consisting of a system with degenerate many-body spectrum weakly
coupled to metallic leads {\it e.g.} molecular junctions, graphene
or carbon nanotube quantum dots, artificial molecules. In
particular, the algebraic formulation of the blocking condition in
terms of kernels of the tunnelling  matrices $T^{\pm}$,
Eq.~\eqref{eq:Blocking_conditions}, allows a straightforward
numerical implementation and makes the algorithm directly
applicable to complex junctions with highly degenerate spectrum.

\section{Aknowledgments}

We acknowledge financial support by the DFG under the programs
SFB689, SPP1243.

\begin{appendix}

\section{}
\label{app:Equivalence}

We derive here the equation (1) in [\onlinecite{Darau09}] as a
specific example of the general theory presented in the paper.
That equation represents the interference blocking condition for
the simplest possible configuration involving only a non
degenerate and a doubly degenerate state.

Let us consider for simplicity a spinless \cite{Spinless} system
and a gate and bias condition that restricts the set of relevant
many-body states to three: one with $N$ particles and two
(degenerate) with $N+1$ or $N-1$ particles. The interference
blocking state, if it exists, belongs to the $N\pm1$ level. There
is only one interesting tunnelling matrix to be analyzed, namely
$T^\mp_{N\pm1}$. Let us take for it the generic form:
\begin{equation}
T^{\mp}_{N\pm1} = \left(%
\begin{array}{cc}
  \gamma_{S1} & \gamma_{S2} \\
  \gamma_{D1} & \gamma_{D2} \\
\end{array}%
\right) \label{eq:T_simple}
\end{equation}
where $S$ and $D$ indicate source and drain respectively and 1 and
2 label the two degenerate states with $N\pm1$ particles.
$\gamma_{S(D)i}$ are the elements of the $t^{\mp}_{S(D)}$ matrices
introduced in Eqs.~\eqref{eq:t_+} and \eqref{eq:t_-}.

The decoupled space reads:
\begin{equation}
\mathcal{D}_{N\pm1} = {\rm ker}\,T^\mp_{N\pm1}.
 \label{eq:D_simple}
\end{equation}

Since the $N\pm1$ particles relevant Hilbert space has dimension 2
the only possibility to find a blocking state is that
$\mathcal{D}_{N\pm1} = \emptyset$. In other terms:

\begin{equation}
{\rm det}\,T^\mp_{N\pm1} = \gamma_{S1}\gamma_{D2} -
\gamma_{D1}\gamma_{S2}\neq 0
 \label{eq:Condition_simple}
\end{equation}
This condition is identical to Eq. (1) in \cite{Darau09}. The
blocking state can finally be calculated as:
\begin{equation}
 \begin{split}
\mathcal{B}_{N+1} &= \mathcal{P}_{N+1} {\ker} \left(%
\begin{array}{ccc}
  \gamma_{S1} & \gamma_{S2} & 1 \\
  \gamma_{D1} & \gamma_{D2} & 0 \\
\end{array}%
\right) \cap \mathcal{C}_{N+1}\\
\text{or}\\
\mathcal{B}_{N-1} &= \mathcal{P}_{N-1} {\ker} \left(%
\begin{array}{ccc}
  \gamma_{S1} & \gamma_{S2} & 0 \\
  \gamma_{D1} & \gamma_{D2} & 1 \\
\end{array}%
\right) \cap \mathcal{C}_{N-1},
 \end{split}
\end{equation}
where the $\mathcal{C}_{N\pm1}$ is, in the relevant case, the
entire space and the projector $\mathcal{P}_{N\pm1}$ simply
removes the last component of the vector that defines the one
dimensional kernel.

\section{}
\label{app:T_matrices}

We give here explicitly the $T^-_{2,2_1\,1_0}$ matrix necessary
for the calculation of the triplet blocking states and the
associated blocking states. The states in the $1_0$  doublet and
in the two times orbitally degenerate triplet $2_1$ are labelled
and ordered as follows:
\begin{equation}
1_0\,\left\{
 \begin{array}{l}
 |1_0, \ell=0,\uparrow\rangle\\
 |1_0, \ell=0,\downarrow\rangle
 \end{array}
 \right.,
 \quad
 2_1\,\left\{
 \begin{array}{l}
   |2_1, \ell=+\hbar, S_z=+\hbar\rangle \\
   |2_1, \ell=+\hbar, S_z=0\rangle \\
   |2_1, \ell=+\hbar, S_z=-\hbar\rangle \\
   |2_1, \ell=-\hbar, S_z=+\hbar\rangle \\
   |2_1, \ell=-\hbar, S_z=0\rangle \\
   |2_1, \ell=-\hbar, S_z=-\hbar\rangle \\
 \end{array}
 \right..
 \label{eq:vectors}
\end{equation}
The elements of the $t^-_{\alpha \sigma}$ matrices that compose
$T^-_{2,2_1\,1_0}$ have thus the general form:
$$t^-_{\alpha \sigma}(S_z,S'_z,\ell) = \langle 1_0,\ell'=0,S_z'|
 d^{\phd}_{\alpha \sigma}
 |2_1,\ell,S_z\rangle.$$
By orbital and spin symmetry arguments it is possible to show that
$$t^-_{\alpha \sigma}(S_z,S'_z,\ell) =
t\,e^{\frac{i}{\hbar}\ell\phi_{\alpha}}\,\delta_{S'_z,S_z-\sigma}\,(\sqrt{2}\delta_{S'_z,\uparrow}.
+ \delta_{S'_z,\downarrow})$$
where $$t = \langle 1_0,\ell'=0,\downarrow|
 d^{\phd}_{M \uparrow}
 |2_1,\ell = 1,S_z = 0 \rangle.$$
The subscript $M$ labels a reference dot and $\phi_{\alpha}$ is
the angle of the rotation that brings the dot $\alpha$ on the dot
$M$. The explicit form of $T^-_{2_1,2_1\,1_0}$ reads:
\begin{widetext}
 \begin{equation}
    T^-_{2,2_1\,1_0} = t
    \left(%
    \begin{array}{cccccc}
      \sqrt{2}e^{-i2\pi/3} & 0 & 0 & \sqrt{2}e^{+i2\pi/3} & 0 & 0 \\
      0 & e^{-i2\pi/3} & 0 & 0 & e^{+i2\pi/3} & 0 \\
      0 & e^{-i2\pi/3} & 0 & 0 & e^{+i2\pi/3} & 0 \\
      0 & 0 & \sqrt{2}e^{-i2\pi/3} & 0 & 0 & \sqrt{2}e^{+i2\pi/3} \\
      \sqrt{2}e^{+i2\pi/3} & 0 & 0 & \sqrt{2}e^{-i2\pi/3} & 0 & 0 \\
      0 & e^{+i2\pi/3} & 0 & 0 & e^{-i2\pi/3} & 0 \\
      0 & e^{+i2\pi/3} & 0 & 0 & e^{-i2\pi/3} & 0 \\
      0 & 0 & \sqrt{2}e^{+i2\pi/3} & 0 & 0 & \sqrt{2}e^{-i2\pi/3} \\
    \end{array}%
    \right).
 \end{equation}
\end{widetext}

The rank of this matrix is 6 since all columns are independent.
Thus $\mathcal{C}_{2,2_1}$ coincides with the full Hilbert space
of the first excited 2 electron energy level. The blocking space
$\mathcal{B}_{2,2_1,1_0}$ reads:
\begin{equation}
 \mathcal{B}_{2,2_1,1_0} = \mathcal{P}_{2_1}{\rm ker}(T^-_{2,2_1\,1_0}, T_S)
\end{equation}
where $T_S$ reads

\begin{equation}
 T_S = \left(%
\begin{array}{cccccccc}
  1 & 0 & 0 & 0 & 0 & 0 & 0 & 0 \\
  0 & 1 & 0 & 0 & 0 & 0 & 0 & 0 \\
  0 & 0 & 1 & 0 & 0 & 0 & 0 & 0 \\
  0 & 0 & 0 & 1 & 0 & 0 & 0 & 0 \\
\end{array}%
\right)^T,
\end{equation}
in accordance to its general definition given in
Eq.~\eqref{eq:Blocking_space}, and the projector
$\mathcal{P}_{2_1}$ removes the last four components from the
vectors that span ${\rm ker}(T^-_{2,2_1\,1_0},T_S)$. It is then
straightforward to calculate the vectors that span the blocking
space $\mathcal{B}_{2,2_1,1_0}$:

\begin{equation}
 \mathbf{v}_1 = \left(%
\begin{array}{c}
  \frac{e^{-i\frac{\pi}{6}}}{\sqrt{2}} \\
  0 \\
  0 \\
  \frac{e^{+i\frac{\pi}{6}}}{\sqrt{2}} \\
  0 \\
  0 \\
\end{array}%
\right),
\quad
\mathbf{v}_2 = \left(%
\begin{array}{c}
  0 \\
  \frac{e^{-i\frac{\pi}{6}}}{\sqrt{2}} \\
  0 \\
  0 \\
  \frac{e^{+i\frac{\pi}{6}}}{\sqrt{2}} \\
  0 \\
\end{array}%
\right),
\quad
\mathbf{v}_3 = \left(%
\begin{array}{c}
  0 \\
  0 \\
  \frac{e^{-i\frac{\pi}{6}}}{\sqrt{2}} \\
  0 \\
  0 \\
  \frac{e^{+i\frac{\pi}{6}}}{\sqrt{2}} \\
\end{array}%
\right).
\end{equation}

The vectors $\mathbf{v}_1,\,\mathbf{v}_2$ and $\mathbf{v}_3$ are
the components of the blocking states written in the $2_1$ basis
set presented in \eqref{eq:vectors}. Thus, the three blocking
states correspond to the three different projectors of the total
spin $S_z = \hbar,\,0,\, \text{and} -\hbar$, respectively.

\section{}
\label{app:Effective Hamiltonian}

We present here explicitly the renormalization frequency
$\omega_{\alpha S_z}$ and the projection of the angular momentum
$L_{\alpha}$ which appear in the expression of the effective
Hamiltonian \eqref{eq:H_eff}. The frequency $\omega_{\alpha S_z}$
is defined for the degenerate two particle excited level $2_1$ in
terms of transition amplitudes to all the states of neighbor
particle numbers:

\begin{equation}
\begin{split}
 \omega_{\alpha S_z} &= \frac{1}{\pi}\sum_{\sigma' E}
\Gamma^0_{\alpha\sigma'}\\
 \Big[
 & \langle 2_1\ell S_z|
 d^{\phd}_{M \sigma'}\,\mathcal{P}_{3E}\,d^\dagger_{M \sigma'}
 | 2_1 \,-\!\ell \, S_z \rangle
 p_\alpha(E - E_{2_1}) +\\
 & \langle 2_1\ell S_z|
 d^\dagger_{M \sigma'}\,\mathcal{P}_{1E}\,d^{\phd}_{M \sigma'}
 | 2_1 \,-\!\ell\, S_z \rangle
 p_\alpha(E_{2_1} \!\!- E)\Big],
\end{split}
\label{eq:frequencies}
\end{equation}
%
where $\mathcal{P}_{NE} \equiv \sum_{m \tau}|N m \tau E \ixe N m
\tau E|$ is the projector on the $N$-particle level with energy
$E$ and $d^{\phd}_{{\rm M} \sigma}$ destroys an electron of spin
$\sigma$ in the middle dot $M$. We defined the function
$p_\alpha(x)=-{\rm Re}\psi\left[\tfrac{1}{2} +
\tfrac{i\beta}{2\pi}(x -\mu_\alpha)\right]$, where $\beta =
1/k_{\rm B}T$, $T$ is the temperature  and $\psi$ is the digamma
function. Moreover $\Gamma^0_{\alpha\sigma'} =
\tfrac{2\pi}{\hbar}|t|^2 D_{\alpha\sigma'}$ is the bare tunnelling
rate to the lead $\alpha$ of an electron of spin $\sigma'$, where
$t$ is the tunnelling amplitude and $D_{\alpha\sigma'}$ is the
density of states for electrons of spin $\sigma'$ in the lead
$\alpha$ at the corresponding chemical potential $\mu_{\alpha}$.
Due to the particular choice of the arbitrary phase of the $2$
particle excited states, $\omega_{\alpha S_z}$ does not depend on
the orbital quantum number $\ell$. It depends instead on the bias
and gate voltage through the energy of the $1$, $2_1$ and $3$
particle states.

In the Hilbert space generated by the two-fold orbitally
degenerate $|2_1\,\ell\,S_z\rangle$ the operator $L_{\alpha}$
reads:

\begin{equation}
L_{\alpha}= \frac{\hbar}{2} \left( \begin{array}{cc}
1 & e^{i2\phi_{\alpha}}  \\
e^{-i2\phi_{\alpha}} & 1 \\
 \end{array} \right),
\label{eq:angular_momentum}
\end{equation}
where $\phi_{\alpha} = \pm \tfrac{2\pi}{3}$ is the angle of which
we have to rotate the triple dot system to bring the middle dot
$M$ into the position of the contact dot $\alpha$. For a
derivation of (\ref{eq:angular_momentum}) see the supplementary
material of [\onlinecite{Donarini09}]. For all degenerate
subspaces, if no accidental degeneracy is present (like for our
parameter choice), the effective Hamiltonian has the form given in
\eqref{eq:H_eff}, \eqref{eq:frequencies},
\eqref{eq:angular_momentum}, with the renormalization frequencies
calculated using the appropriate energies and matrix elements.
\end{appendix}


\begin{thebibliography}{99}
\bibitem{Free_energy} If the equilibrium chemical potential is not
set to zero the many-body energy spectrum should be substituted
with the spectrum of the many-body free energy ($H - \mu_0 N$)
where $\mu_0$ is the chemical potential of the leads at zero bias.
The rest of the argumentation remains unchanged.
\bibitem{Spinless} The assumption of a spinless system is not
restrictive for parallel polarized leads and transitions between a
spin singlet and a doublet since the different spin sectors
decouple from each other.
\end{thebibliography}
\end{document}